# Saturation Throughput - Delay Analysis of IEEE 802.11 DCF in Fading Channel

Zoran Hadzi-Velkov

Ss. Cyril and Methodius University
Faculty of Electrical Engineering
1000 Skopje, Macedonia
zoranhv@etf.ukim.edu.mk

Boris Spasenovski

Ss. Cyril and Methodius University
Faculty of Electrical Engineering
1000 Skopje, Macedonia
boriss@etf.ukim.edu.mk

*Abstract*– **In this paper, we analytically analyzed the impact of an error-prone channel over all performance measures in a traffic-saturated IEEE 802.11 WLAN. We calculated station's transmission probability by using the modified Markov chain model of the backoff window size that considers the frame-error rates and maximal allowable number of retransmission attempts. The frame error rate has a significant impact over theoretical throughput, mean frame delay, and discard probability. The peak throughput of a WLAN is insensitive of the maximal number of retransmissions. Discard probabilities are insensitive to the station access method, Basic or RTS/CTS.**

## I. INTRODUCTION

In recent years, IEEE 802.11 networks are becoming a predominant technology for wireless connectivity in local areas. IEEE 802.11 standard [1] has been developed to provide high bandwidth to mobile users in indoor environments. However, the radio channel introduces significant complexity to the design and performance analysis of the WLANs. This is primarily due to multipath fading, which produces high error rates, depending on the channel conditions, signal rates, and station mobility. The IEEE 802.11 performance, primarily the throughput, has been studied in number of papers both analytically [2-8] and by simulation, but none of them consider mean frame delays and frame-loss ratios based on an analytical model. Further, up to author's knowledge the impact of frame-error rates also has not be considered analytically.

In this paper, we provide an analysis of all performance measures of the IEEE 802.11 Distributed Coordination Function (DCF) in saturation in non-ideal channel conditions. By extending the Markov chain model from [7], we were able to produce analytical solutions for the peak system capacity, mean frame delays, and discard probabilities in a saturated WLAN exposed to an error-prone radio channel.

## II. PERFORMANCE ANALYSIS

### A. Modified Markov chain model

For the purpose of the analysis, we need to determine the transmission probability $\tau$ of each station in a randomly chosen slot time. Thus, we used the discrete Markov chain model from [7], which relates only to stations with "persist-until-success" retransmission strategy in ideal channel conditions. In this paper, we extend this model by taking the frame-error probability $P_f$ into account. Additionally, we consider the finite number of retransmission attempts ($m + f + 1$) after which the frame is discarded from the transmit queue and a new frame is admitted in the queue.

Let us consider finite number of stations in the network, $n$. In saturating conditions, after successful transmission or discard of a frame, each station has immediately a new frame available for transmission, i.e. its queue is always assumed to be non-empty. Under such conditions, it is reasonable to assume that after performing *Carrier Sensing*, the station will find the channel occupied, re-enter backlog condition, and immediately start executing *Collision Avoidance* procedure, i.e. the binary exponential backoff algorithm. Thus, starting with the very first transmission attempt, the station tries to access the channel after performing random backoff. The finite-state model of each station is represented by the two-dimensional Markov chain of its backoff window size, which is depicted in Fig. 1.

A current state ($i$, $k$) of a station is determined by the current value of its backoff timer $k \in (0, W_i - 1)$ after it suffered $i$ previous unsuccessful transmission attempts (row $i$ in Fig. 1). Starting with the very first transmission attempt (backoff stage $i = 0$), the initial value of the backoff timer is uniformly chosen in the range between 0 and $W_0 - 1$. After the station enters backoff stage $i$, its backoff timer is reinitialized to a random value between 0 and $W_i - 1$ (slots). After $m + f + 1$ unsuccessful retransmission attempts, the frame is dropped from queue. Until the $m$-th retransmission attempt, the maximal backoff timer $W_i$ increases by factor of 2, after which it is frozen to $W_m$ until the $m + f + 1$ retransmission when the frame is successfully transmitted or discarded, i.e.

$$W_i = \begin{cases} W \cdot 2^i, & 0 \le i \le m-1 \\ W \cdot 2^m, & m \le i \le m+f \end{cases}, \qquad (1)$$

where $W$ is the initial contention window.

The backoff timer is decremented by 1 in each consecutive slot. However, the slot duration differs: if it is an idle slot - the slot lasts $\sigma$ = 20µs; if the slot is occupied - it can be a successful transmission slot, an unsuccessful transmission slot due to frame error, or a collision slot. In the beginning of a busy slot, each backlogged station decrements its timer by one and then the timer is freezed until the channel becomes idle.

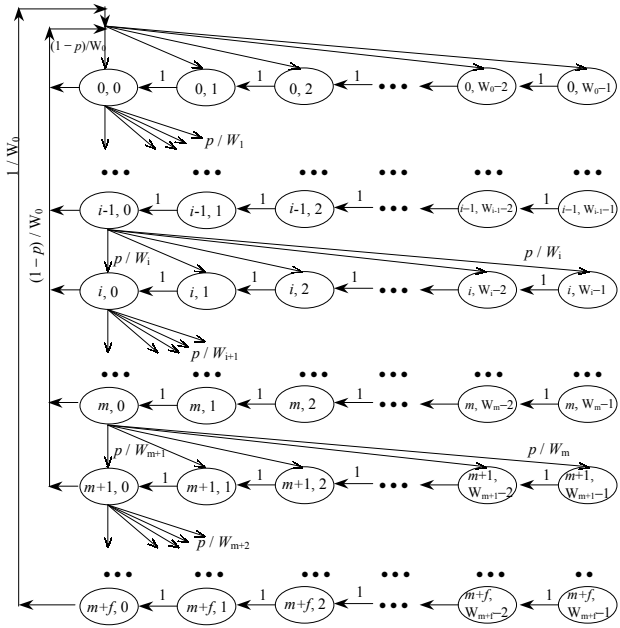

Figure 1. Finite-state station model in saturation based on Markov chain of the backoff window size

We denote the transition probability from one stage to another (e.g. from row $i - 1$ to row $i$ in Fig. 1) by $p$. It is also the probability of an unsuccessful (re)transmission attempt seen by a test station as its frame is being transmitted on the channel. The unsuccessful (re)transmission attempt can happen due: collision of this station with at least one of the $n - 1$ remaining stations, occurring with probability $p_1$,

$$p_1 = 1 - (1 - \tau)^{n-1}, \quad (2)$$

*and/or* an errored frame, occurring with probability $P_f$ (due to the channel fading and/or noise). Since both events are independent, the probability $p$ can be expressed as:

$$p = 1 - (1 - p_1)(1 - P_f) = p_1 + P_f - p_1 P_f . \quad (3)$$

In case of an unsuccessful transmission attempt, after backoff timer expiry in state $(i - 1, 0)$, the station moves in any state on row $i$ $(i, k)$ with probability $p/W_i$. Following a successful transmission (occurring with conditional probability $1 - p$) while the observed station is in stage $i \in (0, m + f - 1)$, a new packet is admitted in the queue, the station returns in backoff stage 0, and its backoff timer uniformly selects any integer value in the range $(0, W_0 - 1)$ with probability $(1 - p)/W_0$. If the station reaches backoff stage $m + f$, and once its backoff timer reaches 0, its frame can be successfully or unsuccessfully transmitted. In both cases, a new frame is admitted in the queue and the station returns in backoff stage 0, and its backoff timer is uniformly chosen in the range $(0, W_0 - 1)$ with probability $1/W_0$.

Let the *stationary* distribution of the chain be $b_{i,k}$, denoting the probability of the station to be in state $(i, k)$. The probability of station to be in state $(i, 0)$ can be expressed through the probability of station to be in state $(i - 1, 0)$ as follows:

$$b_{i,0} = b_{i-1,0} \cdot p, \quad 0 < i \leq m + f \quad (4)$$

which yields to:

$$b_{i,0} = p^i \cdot b_{0,0}, \quad 0 < i \leq m + f \quad (5)$$

Since transmission occurs only in states $(i, 0)$, the probability $\tau$ of a station transmitting in a randomly chosen slot can be expressed as:

$$\tau = \sum_{i=0}^{m+f} b_{i,0} = \sum_{i=0}^{m+f} p^i \cdot b_{0,0} = \frac{1 - p^{m+f+1}}{1 - p} \cdot b_{0,0}. \quad (6)$$

Then, probability $b_{i,k}$ for $0 < i \leq (m+f)$, can be given simply as

$$b_{i,k} = \frac{W_i - k}{W_i} p \cdot b_{i-1,0} = \frac{W_i - k}{W_i} b_{i,0}, \quad 0 < i \leq (m + f), \quad (7)$$

and $b_{0,k}$ as

$$b_{0,k} = (1 - p) \frac{W_0 - k}{W_0} \sum_{j=0}^{m+f-1} b_{j,0} + \frac{W_0 - k}{W_0} b_{m+f,0}$$

$$= (1 - p) \frac{W_0 - k}{W_0} \sum_{j=0}^{m+f-1} p^j b_{0,0} + \frac{W_0 - k}{W_0} p^{m+f} b_{0,0}$$

$$= \frac{W_0 - k}{W_0} b_{0,0} . \quad (8)$$

From (7) and (8), $b_{i,k}$ can generally be expressed:

$$b_{i,k} = \frac{W_i - k}{W_i} b_{i,0}, \quad 0 \leq i \leq (m + f) . \quad (9)$$

Normalizing the stationary distribution of the chain to 1, and using (1) and (9), we have:

$$1 = \sum_{i=0}^{m+f} \sum_{k=0}^{W_i - 1} b_{i,k} = \sum_{i=0}^{m+f} b_{i,0} \sum_{k=0}^{W_i - 1} \frac{W_i - k}{W_i} = \sum_{i=0}^{m+f} b_{i,0} \cdot \frac{W_i + 1}{2}$$

$$= \frac{b_{0,0}}{2} \left[ \sum_{i=0}^{m} p^i (W \cdot 2^i + 1) + \sum_{i=m+1}^{m+f} p^i (W \cdot 2^m + 1) \right] = \frac{b_{0,0}}{2} \cdot \frac{1 - p^{m+f+1}}{1 - p}$$

$$+ W \frac{b_{0,0}}{2} \cdot \frac{[1 - (2p)^{m+1}](1 - p) + p(2p)^m (1 - p^f)(1 - 2p)}{(1 - p^{m+f+1})(1 - 2p)} \quad (10)$$

Finally, we attain the probability $\tau$:

$$\tau = \frac{2(1 - 2p)(1 - p^{m+f+1})}{(1 - 2p)(1 - p^{m+f+1}) + W[1 - p - p(2p)^m (1 + p^f - 2p^{1+f})]} \quad (11)$$

Equations (3) and (11) represent a non-linear system with single solution, which we solve using *Mathematica*. It is obvious that there is a single solution of $\tau$ for each $n$, $W$, $m$, $f$, and $P_f$, i.e. $\tau = f(n, W, m, f, P_f)$. The model from Fig. 1 and (11) are generalizations of the model from [7]. Actually, one can obtain the corresponding results from [7] by solving the non-linear system (3)–(11) in the special case for $f \to \infty$ and $P_f = 0$.

## B. Saturation Throughput

Following a similar reasoning from [7], we can express the normalized saturation throughput of IEEE 802.11 DCF within a single WLAN cell in an error-prone channel as follows:

$$S_{max} = \frac{P_s P_{tr}(1-P_f) \cdot E[L]}{(1-P_{tr})\sigma + P_{tr}P_s(1-P_f)T_s + P_{tr}(1-P_s)T_c + P_{tr}P_s P_f T_e} \quad (12)$$

In (12), $E[L]$ is the average frame payload size, although to establish upper performance limit in the numerical analysis, we assumed all generated packets are fixed and maximized so that $E[L] = L = 2312$ octets. $P_{tr}$ is the probability of at least one transmission in the observed time slot, $P_{tr} = 1 - (1 - \tau)^n$. Thus, the probability of an empty slot is $1-P_{tr}$. $P_s$ is the probability of a single successful transmission given at least one station (out of $n$ stations) is transmitting, $P_s = n\tau(1-\tau)^{n-1}/P_{tr}$. The probability of successful transmission in a slot time is denoted by $P_{tr}P_s(1-P_f)$, the unsuccessful transmission probability due to simultaneous transmission in the same slot (i.e. collision) is $P_{tr}(1-P_s)$, and the unsuccessful transmission probability due to errored frame is $P_{tr}P_s P_f$.

$T_s$ is the average time the channel is sensed busy by each station because of a successful transmission, $T_c$ is the average time the channel is sensed busy during a collision, and $T_e$ is the average time it is sensed busy from a frame which suffered transmission errors. Assuming the duration announcements (contained in the preamble/header part of the frame) are always successfully received by all stations, and the frame errors can occur only in the remaining part of the frame, it is clear that $T_e = T_s$. The values of $T_s$ and $T_c$ differ depending on the network access mode (given below for Basic and RTS/CTS) and additional network operating parameters (Table I)

TABLE I. RELEVANT NETWORK PARAMETERS

| Parameter | Default |
|---|---|
| Channel Rate | 11 Mbps |
| PHY Preamble | 144 symbols |
| PHY Header | 48 symbols |
| MAC header | 34 octets |
| ACK | 14 octets + $PHY_{pre/hdr}$ |
| RTS | 20 octets + $PHY_{pre/hdr}$ |
| CTS | 14 octets + $PHY_{pre/hdr}$ |
| SIFS | 10 μs |
| DIFS | 50 μs |
| Slot_Time σ | 20 μs |
| $m$ | 5 |
| Initial contention window $W$ | 8 |
| $T_s^{basic}$ | 2160.4 μs |
| $T_c^{basic}$ | 1948.2 μs |
| $T_s^{rts/cts}$ | 2589.1 μs |
| $T_c^{rts/cts}$ | 256.5 μs |

Basic: $\begin{cases} T_s^{bas} = PHY_{pre/hdr} + MAC_{hdr} + L + SIFS + ACK + DIFS \\ T_c^{bas} = PHY_{pre/hdr} + MAC_{hdr} + L + DIFS \end{cases}$

RTS/CTS: $\begin{cases} T_s^{rts/cts} = RTS + SIFS + CTS + SIFS + PHY_{pre/hdr} \\ \qquad\qquad + MAC_{hdr} + L + SIFS + ACK + DIFS \\ T_c^{rts/cts} = RTS + DIFS \end{cases}$

## C. Frame Discard Probability

The probability of a frame discard $P_D$ is actually the probability of occurrence of consecutive $m + f + 1$ unsuccessful retransmission attempts, after which the frame is discarded, a new frame is admitted in the transmit queue, and the station returns to backoff stage $i = 0$. Thus,

$$P_D = p^{m+f+1} = [1-(1-P_f)(1-\tau)^{n-1}]^{m+f+1} . \quad (13)$$

Note that discard probabilities are insensitive to the utilized station's access method (Basic or RTS/CTS).

## D. Mean Delay

Now let us concentrate on a single station to determine the average delay $T_d$ of each frame from the moment the backoff procedure is initiated until frame's successful transmission. During the backoff defer slots of the observed station in the $i$-th stage, the probability of transmission of at least one of the $n - 1$ remaining stations slots is $p_1$, while the probability of exactly one transmission from one of the $n - 1$ remaining stations (given at least one of them is transmitting) is:

$$p_{1s} = \frac{(n-1)\tau(1-\tau)^{n-2}}{p_1} . \quad (14)$$

In each backoff stage $i \in (0, m+f)$, the initial value of the backoff timer has mean of $(W_i - 1)/2$, so that the average deferral interval before a retransmission attempt is $(W_i - 1)/2$ slots. The average number of consecutive idle slots $n_{idle}$ between two consecutive busy slots of the $n - 1$ remaining stations can be calculated as:

$$n_{idle} = \sum_{i=0}^{\infty} i(1-p_1)^i p_1 = \frac{1}{p_1} - 1. \quad (15)$$

Thus, a single *renewal cycle* between two consecutive transmissions of the $n - 1$ remaining stations includes multiple consecutive idle slots and an occupied slot, i.e. $n_{rc} = n_{idle} + 1 = 1/p_1$ slots. Since each occupied slot can be a successful, a frame-errored, or a collided slot, the average duration of a renewal cycle $T_{rc}$ is

$$T_{rc} = n_{idle}\sigma + p_{1s}(1-P_f)T_s + (1-p_{1s})T_c + p_{1s}P_f T_e . \quad (16)$$

Since the retransmission attempt of observed station in backoff stage $i$ is on the average preceded by $(W_i - 1)/2 \cdot n_{rc} = (W_i - 1)p_1/2$ renewal cycles of $n - 1$ remaining stations, the average time between two consecutive retransmissions of the observed station is $(W_i - 1)pT_{rc}/2$. The average elapsed time $T_{tct,i}$ before the test

station makes its ($i + 1$)-th retransmission attempt (row $i$ in Fig. 1), can be calculated as:

$$T_{tct,i} = \sum_{k=0}^{i}\left(\frac{W_k - 1}{2} p_1 T_{rc}\right) + i T_{coe}$$
$$= \frac{p_1 T_{rc}}{2} \sum_{k=0}^{i} W_k - (i+1)\frac{p_1 T_{rc}}{2} + i T_{coe} \quad . \quad (17)$$

where $T_{coe}$ is the average duration while the observed station itself occupies the channel during each unsuccessful retransmission attempt:

$$T_{coe} = \frac{p_1(1-P_f)T_c + p_1 P_f T_c + (1-p_1)P_f T_e}{p}$$
$$= \frac{p_1 T_c + (1-p_1)P_f T_e}{p} \quad . \quad (18)$$

After substitution of (1) into (17), we have:

$$T_{tct,i} = i T_{coe} - (i+1)\frac{p_1 T_{rc}}{2}$$
$$+ \frac{p_1 T_{rc}}{2} W \cdot \begin{cases} 2^{i+1} - 1, & 0 \le i \le m-1 \\ 2^{m+1} - 1 + 2^m(i-m), & m \le i \le m+f \end{cases} \quad . \quad (19)$$

Finally, the average frame delay until successful transmission is:

$$T_d = \sum_{i=0}^{m+f}(1-p)\cdot p^i \cdot (T_{tct,i} + T_s) \quad . \quad (20)$$

By introducing (19) into (20), the latter can easily be solved in closed-form. In the special case $f \to \infty$, (20) attains a more simpler form:

$$T_d = T_s + (T_{coe} p - \frac{T_{rc} p_1}{2})\frac{1}{1-p} + T_{rc}\frac{W \cdot p_1}{2}\frac{1-p-p(2p)^m}{(1-p)(1-2p)}$$
$$(21)$$

Additionally, to calculate $T_d$ for $f \to \infty$ and $P_f = 0$, one needs to substitute $p_1 = p$ and $T_{coe} = T_c$ into (21).

III. NUMERICAL RESULTS

The throughput and delay performance vs. $n$ for special case $f \to \infty$ and $P_f = 0$ is depicted in Fig. 2. As expected, the saturation throughput decreases with the increase of number of contending stations in Basic access mode, while it remains stable in RTS/CTS access mode. The mean delay increases in both access modes, although it is lower in RTS/CTS access mode since only the short RTS frames participate into the collisions. However, as compared to the ratio between the length of the useful frame and the RTS frame, the difference is not as excessive as one would expect.

The reason is that the frame delay in each station originates from the backofff defer periods in a significant portion, as well as from the periods when the station participates in collisions. Obviously, due to the "persistent-until-success" retransmission strategy ($f \to \infty$), there are no frame losses at all, i.e. $P_D = 0$. We emphasize the graphs from Fig. 2 and all the following results refer to 11 Mbps transmission rate (Table I), while corresponding system parameters must be used according to IEEE 802.11b standard for rates of 1, 2, and 5.5 Mbps. Additionally, the initial contention window is 8 ($W = 8$), while the retransmission attempt threshold (after which the initial backoff window is frozen) is set to 5, i.e. $m = 5$.

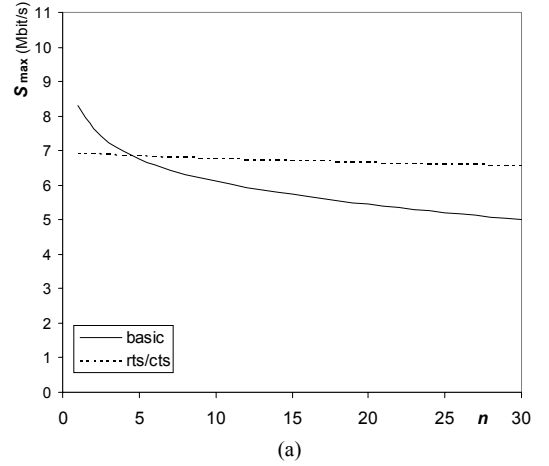

(a)

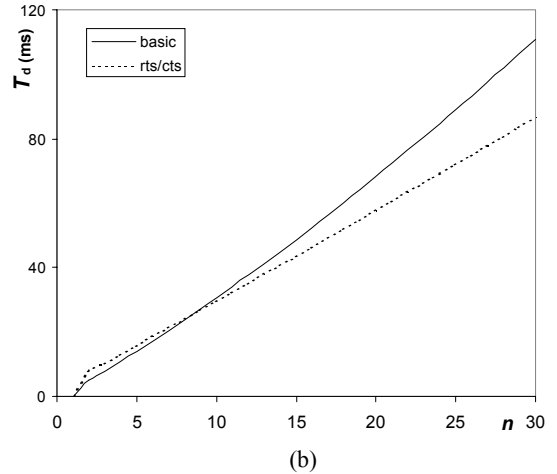

(b)

Figure 2. Network performance in "persist-until-success" retransmission strategy and an error-free channel;
(a) throughput, (b) delay

The maximal allowable number of retransmission attempts ($m + f + 1$) has minor impact over the maximum achievable throughput $S_{max}$. However, its influence over mean delay $T_d$ and discard probability $P_D$ cannot be disregarded.

Fig. 3 displays the mean frame delay and discard probabilities in function of $f$, with $P_f = 0.1$ and 0.5 as curve parameters, and $n = 30$. It is reasonable to expect that mean delay $T_d$ will increase (Fig. 3a) and the frame losses $P_D$ would decrease (Fig. 3b) as we allow for higher number of retransmission attempts. In Fig. 3b, the curves for Basic and RTS/CTS access modes for given $P_f$ coincide.

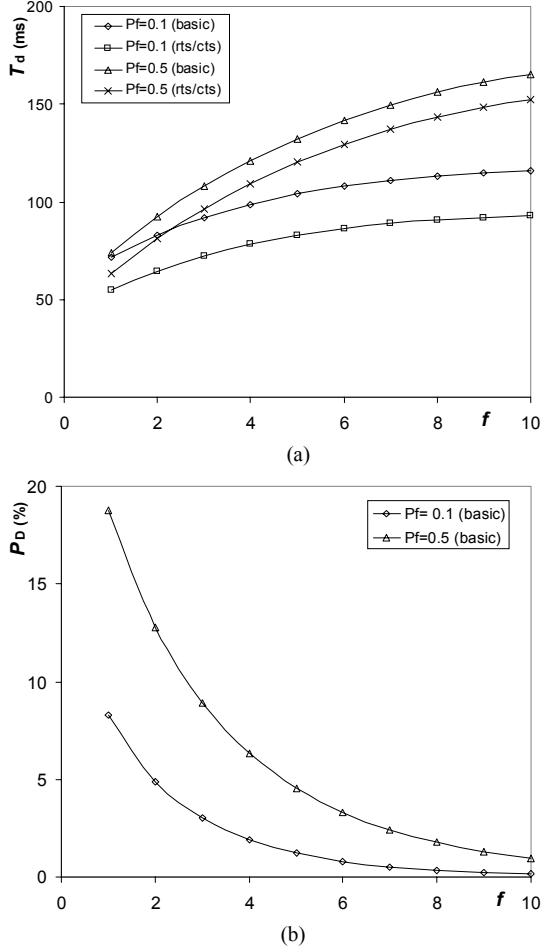

Figure 3. Influence of retransmission threshold $f$ over performance: (a) delay, (b) discard probability

The impact of frame-error rates $P_f$ over all performance measures is depicted in Fig. 4; $f$ appears as parameters in the graphs ($f = 1$ and $f = 10$), while $n = 30$. Increasing $P_f$ from 0.01 to 1, throughput degrades towards 0 (Fig. 4a), and discard probability increases towards 100 % (Fig. 4c). Saturation mean delay also increases with the increase of $P_f$, however as $P_f$ approaches near 1, the mean delay sharply decreases to 0 because almost every frame is discarded (Fig. 4b).

At this point, we note that the graphs of $P_f$ vs. channel signal-to-noise ratio SNR in [9] can be used to calculate the corresponding curves of the three performance measures for the IEEE 802.11b WLAN in function of both the SNR and the delay spread of the multipath-faded channel.

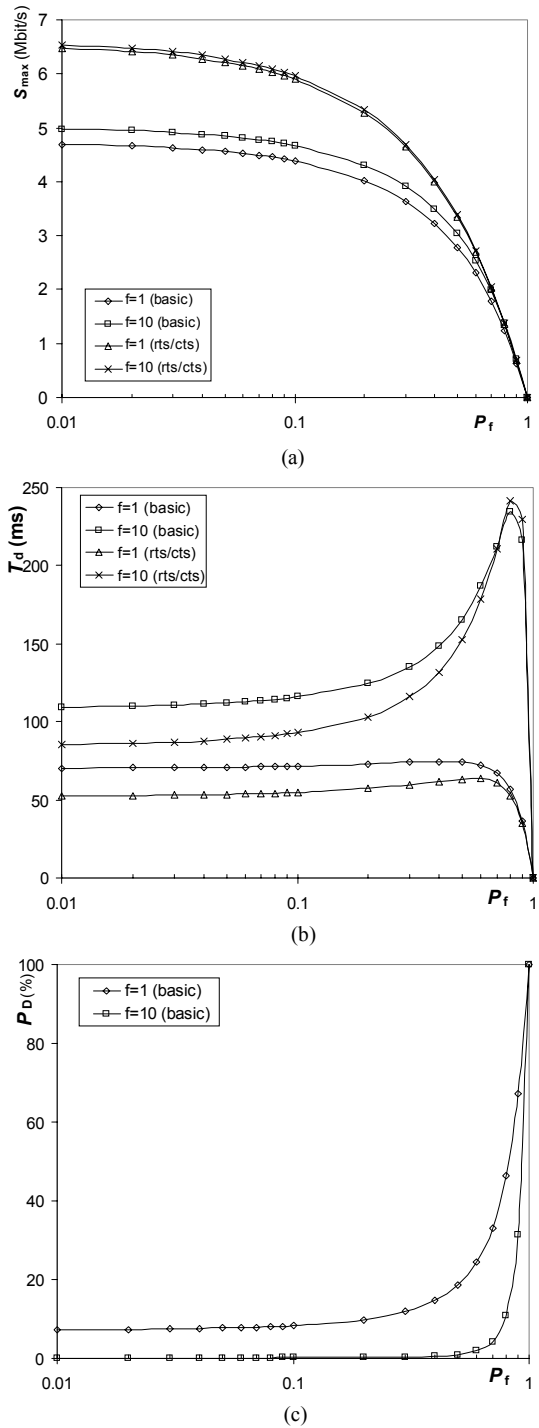

Figure 4. Impact of frame-error rate $P_f$ over performance: (a) saturation throughput, (b) delay, (c) discard probability

## IV. Conclusion

Based on our modified discrete Markov model for the backoff window size with finite number of retransmission attempts, we calculated the transmission probability in a slot time of each IEEE 802.11 station operating in saturated conditions in an error-prone radio channel. The transmission probability proves to depend on the number of contending stations, the frame error probability, and the network operating parameters. Using the transmission probability, we provide analytical solutions for all relevant performance measures of a single IEEE 802.11 cell: network throughput, and station's mean delay and discard probability of the DCF in saturation for both access modes, Basic and RTS/CTS. Increasing the maximal number of retransmissions after which a frame is discarded from station's queue, the delay increases and discard probability decreases, while the saturation throughput is practically unchanged for both access modes. Conversely, the frame-error rate $P_f$ significantly impacts only the throughput for $0.01 < P_f < 0.1$, while for high error rates - all performance measures are affected.

Finally, let's emphasize that in this paper we only considered the impact of the frame-error rate. The multipath fading channel introduces additional complexity affecting wireless networks performance, e.g. the capture effect. Refer to [8] for more detailed saturation throughput analysis of IEEE 802.11b DCF under capture